\documentclass[preprint,showpacs,nofootinbib,aps]{revtex4}
 \usepackage{graphicx}
 \usepackage{bm}
 
\begin{document}
 \title{Study of $B_{c}$ ${\to}$ $KK$ decay with perturbative QCD approach}
 \author{Yueling Yang}
 \affiliation{College of Physics and Information Engineering,
              Henan Normal University,
              Xinxiang 453007, China}
 \author{Junfeng Sun}
 \thanks{corresponding author}
 \affiliation{College of Physics and Information Engineering,
              Henan Normal University,
              Xinxiang 453007, China}
 \author{Na Wang}
 \affiliation{College of Physics and Information Engineering,
              Henan Normal University,
              Xinxiang 453007, China}

 \begin{abstract}
 In the framework of the perturbative QCD approach, we study
 the charmless pure weak annihilation $B_{c}^{-}$ ${\to}$
 $K^{-}K^{0}$ decay and find that the branching ratio
 ${\cal BR}(B_{c}{\to}KK)$ ${\sim}$ ${\cal O}(10^{-7})$.
 This prediction is so tiny that the $B_{c}$ ${\to}$ $KK$
 decay might be unmeasurable at the Large Hadron Collider.
 \end{abstract}
 \pacs{12.39.St  13.25.Hw}
 \maketitle

 \section{Introduction}
 \label{sec1}
 The study of the decays of $B$ mesons is important and interesting
 for the determination of the flavor parameters of the Standard Model
 (SM), the exploration of $CP$ violation, the search of new physics
 beyond SM, etc. In recent years, theoretical studies of $B_{u,d}$
 mesons have been investigated widely in the literatures. They are
 tested and supported by the experimental data collected by the detectors
 at the $e^{+}e^{-}$ colliders, such as the CLEO, Babar, and Belle.
 With the bright hope arising from the startup of the CERN
 Large Hadron Collider (LHC) \cite{epjc65p111}, the heavier $B_{s}$ and
 $B_{c}$ mesons could be produced abundantly and studied in detail at
 the hadron colliders.
 It is estimated that one could expect around $5$ ${\times}$ $10^{10}$
 ``self-tagging'' $B_{c}$ events per year at the LHC \cite{0412158} due to
 a relatively large production cross section \cite{prd71p074012} plus the
 huge luminosity ${\cal L}$ $=$ $10^{34}$ $\hbox{cm}^{-2}\hbox{s}^{-1}$ \cite{lhc}.
 There seems to exist a real possibility to study not only some $B_{c}$
 rare decays, but also $CP$ violation and polarization asymmetries.
 The study of the $B_{c}$ mesons will highlight the advantages of $B$
 physics.

 The $B_{c}$ mesons are the ``double heavy-flavored'' binding systems
 and share many features with the heavy quarkonia.
 The first observation of the $B_{c}$ mesons at the Tevatron \cite{CDF1998}
 provokes the physicist's particular interest in them. Many studies
 and investigation of the properties of the $B_{c}$ mesons have been
 made, and will be further scrutinized by the LHC experiments.
 Because the $B_{c}$ mesons lie below the $BD$ threshold (here we only
 discuss the lightest $1^{1}S_{0}$ ground state pseudoscalar $B_{c}$
 mesons, excluding their excited states) and carry flavors, they
 cannot annihilate into gluon and/or photon so are stable for the
 strong and/or electromagnetic interaction.
 Because of the flavor quantum numbers $B$ $=$ $-C$ $=$ ${\pm}1$,
 the $B_{c}$ mesons can decay through the weak interaction only.
 The $B_{c}$ mesons have more decay modes than the $B_{u,d,s}$ mesons
 due to several reasons.
 One is that many decay modes, such as $B_{c}$ ${\to}$
 $B_{u,d,s}$ $+$ $X$, are only accessible by $B_{c}$ mesons because of
 their sufficiently large masses. Another is that the $B_{c}$ mesons carry
 open flavors, so either $b$ or $c$ quarks can decay individually.
 The potential decays of the $B_{c}$ mesons permit us to over-constrain
 quantities determined by the $B_{u,d,s}$ meson decays.

 The decays of $B_{c}$ mesons can be divided into three classes:
 (1) the $b$-quark decay (i.e. $b$ ${\to}$ $q_{_{U}}$, where the up-type quark
     $q_{_{U}}$ $=$ $u$, $c$) accompanied with the spectator $c$-quark,
 (2) the $c$-quark decay (i.e. $c$ ${\to}$ $q_{_{D}}$, where the down-type quark
     $q_{_{D}}$ $=$ $d$, $s$) accompanied with the spectator $b$-quark, and
 (3) the annihilation channel (i.e. $B_{c}^{-}$ ${\to}$ ${\ell}^{-}\tilde{\nu}_{\ell}$,
     $q_{_{D}}\bar{q}_{_{U}}$, where the lepton ${\ell}^{-}$ $=$ $e^{-}$,
     ${\mu}^{-}$, ${\tau}^{-}$).
 Among the multitudinous $B_{c}$ decays, the weak annihilation channels are
 expected to take ${\sim}$ $10\%$ shares according to the estimates in
 \cite{0412158,0211021} for which a major part comes from the tree weak
 annihilation process $B_{c}^{-}$ ${\to}$ $s\bar{c}$ which is not
 helicity-suppressed because of the large charm quark mass and
 produces a large weak annihilation branching ratio with charm in
 the final state, while the charmless pure weak annihilation decay
 $B_{c}$ ${\to}$ $KK$ is helicity-suppressed like the $B_{d}$ ${\to}$
 $KK$ decay and would have a very small branching ratio.
 It is highly expected that the LHC experiments might shed light on
 a better understanding of weak annihilation processes for $B_{c}$ mesons.

 In recent years, several attractive methods have been proposed to
 study the nonleptonic $B$ decays,
 such as the QCD factorization (QCDF) \cite{0006124},
 perturbative QCD method (pQCD) \cite{9607214,9701233,0004004},
 soft and collinear effective theory \cite{prd63p114020,prd65p054022}, etc.
 Here, we would like to investigate the charmless pure weak
 annihilation $B_{c}$ ${\to}$ $KK$ decay with the pQCD approach
 due to several reasons.
 (1) One reason is that fits of nonleptonic charmless decays $B_{u,d}$
 ${\to}$ $PP$, $PV$ without taking into account weak annihilation
 contributions are generally of poor quality \cite{npb675p333}
 (here $P$ and $V$ denote the lightest ground pseudoscalar and vector
 mesons, respectively).
 Our present understanding of the weak annihilation contributions
 remains limited and unclear.
 So the pure weak annihilation processes, such as $B_{c}$ ${\to}$
 $KK$ decays, are interesting and worthy of study, which will
 certainly help us to improve our understanding of the weak
 annihilation contributions.
 (2) Another is that due to both kinematic improvement from the large
 phase spaces and dynamic enhancement of the CKM factor
 ${\vert}V_{cb}V_{ud}^{\ast}{\vert}$, the $B_{c}$ ${\to}$ $KK$ decay is
 expected to have a large branching ratio among two-body nonleptonic charmless
 $W$-annihilation $B_{c}$ ${\to}$ $PP$ processes. In addition to the absence
 of penguin operators for the tree annihilation process $B_{c}$ ${\to}$ $KK$,
 final state interactions arising from soft gluon exchanges are expected
 to be extremely small because of the large momenta of the final $K$
 mesons. Therefore a relatively accurate estimation of annihilation
 contributions could be obtained effectively from the charmless
 $B_{c}$ ${\to}$ $KK$ decay.
 (3) Still another is that Ref.\cite{prd65p114007} obtains
 a very large $B_{c}$ ${\to}$ $KK$ branching ratio, about $1.6\%$, at
 $4$ orders of magnitude bigger than the estimate ${\cal O}(10^{-6})$
 of Ref.\cite{prd80p114031}, but this estimate is not valid because
 Ref.\cite{prd65p114007} in their calculation incorrectly uses the
 measured penguin-dominated $B_{u}^{\pm}{\to}{\pi}^{\pm}K$ branching
 ratio while the decay $B_{c}$ ${\to}$ $KK$ is a pure tree weak
 annihilation and should be related to $B_{d}$ ${\to}$ $KK$.
 In addition, the branching ratio of the charmless decay $B_{c}$
 ${\to}$ $KK$ is estimated to be ${\cal O}(10^{-8})$ with the QCDF
 approach \cite{prd80p114031}.
 Recently, this charmless decay is also studied with the pQCD
 approach and its branching ratio is ${\cal O}(10^{-7})$ with the
 off-mass-shell final states \cite{09121163}, which is the same
 order of magnitude as ours obtained in this paper with the on-shell
 final states.

 This paper is organized as follows : In Section \ref{sec2},
 we will discuss the theoretical framework and give the decay amplitudes
 for $B_{c}$ ${\to}$ $KK$ with the perturbative QCD approach.
 In our calculation, we shall ignore the final state interactions
 because the final states have very large momenta and move far away
 before soft gluon exchange.
 Section \ref{sec3} is devoted to the numerical result of the branching
 ratio. Finally, we summarize in Section \ref{sec4}.

 \section{Theoretical framework and the decay amplitudes}
 \label{sec2}
 \subsection{The effective Hamiltonian}
 \label{sec21}
 Using the Operator Product Expansion approach and renormalization
 group (RG) equation, the low-energy effective Hamiltonian for
 $B_{c}$ ${\to}$ $KK$ decay can be written as
 \begin{equation}
 {\cal H}_{eff}\,=\,\frac{G_{F}}{\sqrt{2}} V_{cb}V_{ud}^{\ast} \Big\{
    C_{1}({\mu})Q_{1}+ C_{2}({\mu})Q_{2} \Big\} + \hbox{H.c.},
 \label{eq:hamiltonian}
 \end{equation}
 where $G_{F}$ is the Fermi coupling constant for electroweak interactions.
 $V_{cb}V_{ud}^{\ast}$ is the CKM factor accounting for the strengths of
 the nonleptonic $B_{c}$ decays.
 $C_{i}({\mu})$ are Wilson coefficients at the renormalization scale
 ${\mu}$ which have been evaluated to the next-to-leading order with
 the perturbation theory. The local tree operators are process dependent.
 Their expressions are defined as
 \begin{equation}
 Q_{1}=\big[\bar{c}_{\alpha}{\gamma}_{\mu}(1-{\gamma}_{5})b_{\alpha}\big]
       \big[\bar{d}_{\beta} {\gamma}^{\mu}(1-{\gamma}_{5})u_{\beta} \big],~~~~~
 Q_{2}=\big[\bar{c}_{\alpha}{\gamma}_{\mu}(1-{\gamma}_{5})b_{\beta} \big]
       \big[\bar{d}_{\beta} {\gamma}^{\mu}(1-{\gamma}_{5})u_{\alpha}\big],
 \label{eq:operator}
 \end{equation}
 where ${\alpha}$, ${\beta}$ are $SU(3)$ color indices.
 The most difficult problem in theoretical calculation of nonleptonic
 charmless decay $B_{c}$ ${\to}$ $KK$ is how to evaluate the hadronic
 matrix elements ${\langle}KK{\vert}Q_{1,2}{\vert}B_{c}{\rangle}$
 properly and accurately.

 \subsection{Hadronic matrix elements}
 \label{sec22}
 For convenience, the kinematics variables are described in the terms of
 the light cone coordinate. The momenta of the valence quarks and hadrons
 in the rest frame of the $B_{c}$ meson are defined by
 \[ \begin{array}{lclcl}
    p_{B_{c}^{-}}\,=\,p_{1}\,=\,\frac{m_{_{B_{c}}}}{\sqrt{2}}(1,1,\vec{0}_{\perp}),
    &~~& k_{1}=x_{1}p_{1}+(0,0,\vec{k}_{1{\perp}}), &~~&  \\
    p_{K^{-}}\,=\,p_{2}\,=\,\frac{m_{_{B_{c}}}}{\sqrt{2}}(1,0,\vec{0}_{\perp}), &&
    k_{2}=x_{2}p_{2}+(0,0,\vec{k}_{2{\perp}}), & & n_{2}=(1,0,0), \\
    p_{K^{0}}\,=\,p_{3}\,=\,\frac{m_{_{B_{c}}}}{\sqrt{2}}(0,1,\vec{0}_{\perp}), &&
    k_{3}=x_{3}p_{3}+(0,0,\vec{k}_{3{\perp}}), & & n_{3}=(0,1,0),
    \end{array} \]
 where $n_{2}{\cdot}n_{3}$ $=$ $1$. The null vectors $n_{2}$ and $n_{3}$ are
 the plus and minus directions, respectively. $k_{1}$ is the momentum of $c$
 quark in the $B_{c}$ meson. $k_{2}$ and $k_{3}$ are the momenta of the light
 non-strange quark in the $K^{-}$ and $K^{0}$ mesons, respectively.
 $\vec{k}_{i{\perp}}$ denotes the transverse momentum. $x_{i}$ denotes the
 longitudinal momentum fraction of the valence quark.

 The calculation of the hadronic matrix elements is difficult due to the
 nonperturbative effects arising from the strong interactions.
 Phenomenologically, using the Brodsky-Lepage approach \cite{prd22p2157},
 a modified perturbative QCD formalism has been proposed recently under
 the $k_{T}$ factorization framework \cite{9607214,9701233,0004004}.
 Taking into account the transverse momentum of the valence quarks in
 the hadrons, the Sudakov factors are introduced to modify the endpoint
 behavior of the hadronic matrix elements. The amplitudes are factorized
 into three convolution parts : the ``harder'' functions, the heavy quark
 decay subamplitudes, and the nonperturbative meson wave functions,
 which are characterized by the $W^{\pm}$ boson mass $m_{W}$, the
 typical scale $t$ of the decay processes, and the hadronic scale
 ${\Lambda}_{QCD}$, respectively. The pQCD approach has been
 extensively applied to study semileptonic and nonleptonic $B$ decays
 with phenomenological results. More information about pQCD approach
 can be found in \cite{9607214,9701233,0004004}.
 The final decay amplitudes can be expressed as
 \begin{equation}
 {\cal A}(B_{c}^{-}{\to}K^{-}K^{0})\, {\propto}\,
 C(t){\otimes}H(t){\otimes}{\Phi}_{B_{c}^{-}}(x_{1},b_{1}) {\otimes}
 {\Phi}_{K^{-}}(x_{2},b_{2}){\otimes}{\Phi}_{K^{0}}(x_{3},b_{3}),
 \label{eq:am01}
 \end{equation}
 where the Wilson coefficient $C(t)$ is calculated in perturbative theory
 at the scale of $m_{W}$ and evolved down to the typical scale $t$ using the
 RG equations. ${\otimes}$ denotes the convolution over parton kinematic
 variables. $H(t)$ is the hard-scattering subamplitude which is dominated
 by hard gluon exchange and can be factorized. The universal wave functions
 ${\Phi}(x,b)$ absorb nonperturbative long-distance dynamics, which can be
 extracted from experiments or constrained by lattice calculation and QCD
 sum rules. $b$ is the conjugate variable of the transverse momentum of
 the valence quark of the meson. According to the arguments in
 \cite{9607214,9701233,0004004}, the amplitude of Eq.(\ref{eq:am01}) is
 free from the renormalization scale dependence.

 \subsection{Bilinear operator matrix elements}
 \label{sec23}
 Within the pQCD framework, the long-distance hadronic information is
 contained by the the so-called light-cone distribution amplitudes (LCDAs)
 which are defined from hadron-to-vacuum matrix elements of nonlocal
 bilinear operators. Although LCDAs are not calculable in QCD perturbation
 theory, some of their properties are well understood for both light
 and heavy mesons. For example, the LCDAs for the $K$ meson including higher-twist
 contributions are systematically presented in \cite{jhep0605004}.
 In our calculation, we only consider two-particle (valence quarks) twist-2
 and twist-3 LCDAs for $K$ mesons, and neglect contributions from higher
 Fock states. The LCDAs for $K$ mesons are written as
  \begin{equation}
 {\langle}K(p){\vert}\bar{s}_{\alpha}(0)q_{\beta}(z){\vert}0{\rangle}=
  \!\frac{i}{\sqrt{2N_{c}}}{\int}_{0}^{1}\!{\bf d}x\,e^{ixp{\cdot}z}
  \Big\{\!{\gamma}_{5} \not{\!\!p}\,{\phi}_{K}^{a}(x)
       +\!{\gamma}_{5}{\mu}_{K}\! \Big[ {\phi}_{K}^{p}(x)
       -(\not{\!n}_{_{+}}\!\!\not{\!n}_{_{\!-}}\!-1)
 {\phi}_{K}^{t}(x)\Big]\Big\}_{{\beta}{\alpha}}
 \end{equation}
 where $N_{c}$ is the color number.
 The parameter ${\mu}_{K}$ is the chiral factor ${\mu}_{K}$ $=$
 $m_{K}^{2}/(m_{s}+m_{q})$. The null vector $n_{_{+}}$ and $n_{_{-}}$
 are parallel to $p$ and $z$, respectively. The expressions of the
 twist-2 LCDAs ${\phi}_{K}^{a}$ and the twist-3 LCDAs ${\phi}_{K}^{p}$,
 ${\phi}_{K}^{t}$ are collected in Appendix \ref{app01}.

 Unlike the ${\pi}$ and $K$ mesons, our knowledge of the LCDAs for
 $B_{c}$ mesons has been relatively poor until recently (for a recent view,
 see \cite{jhep0804061}), but we know that the $B_{c}$ mesons are
 composed of heavy valence quark both $b$ and $c$. Given $m_{B_{c}}$
 ${\approx}$ $m_{b}$ $+$ $m_{c}$, the $B_{c}$ mesons can be described
 approximately by nonrelativistic dynamics. In this paper, we
 will take
 \begin{equation}
 {\langle}0{\vert}\bar{c}_{\alpha}(z)b_{\beta}(0){\vert}B_{c}^{-}(p_{1}){\rangle}
 =\frac{if_{B_{c}}}{4N_{c}}{\int}{\bf d}x_{1}\,{\rm e}^{-ix_{1}p_{1}{\cdot}z}
  \Big[ \Big(\!\not{\!\!p}_{1}\!+\!m_{B_{c}} \Big) {\gamma}_{5} {\phi}_{B_{c}}(x_{1})
  \Big]_{{\beta}{\alpha}},
 \label{eq:wf-bc-01}
 \end{equation}
 where $f_{B_{c}}$ is the decay constant of the $B_{c}$ meson.
 As the arguments in \cite{jhep0804061}, this simplest form,
 ${\phi}_{B_{c}}(x)$ $=$ ${\delta}(x-m_{c}/m_{B_{c}})$, is the
 two-particle nonrelativistic LCDAs at the tree level where both heavy
 valence quarks just share the total momentum of the $B_{c}$ mesons
 according to their masses. For a rough estimation of the branching
 ratio for $B_{c}$ ${\to}$ $KK$ decay, we will take the simplest
 form as an approximation, and neglect the relativistic corrections
 and contributions from higher Fock states.

 \subsection{The decay amplitudes}
 \label{sec24}
\begin{figure}[ht]
 \includegraphics[angle=0,width=0.95\textwidth,bb=50 680 530 750]{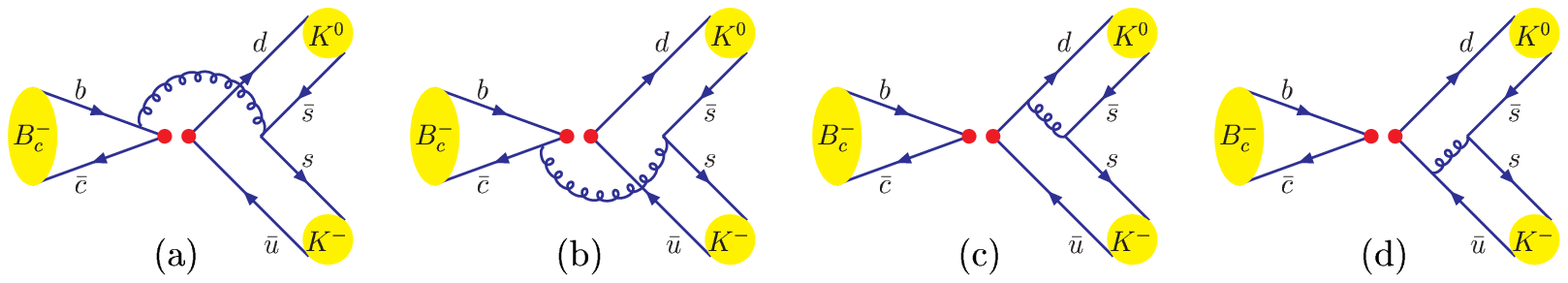}
 \caption{The lower order feynman diagrams contributing to the $B_{c}$
 ${\to}$ $KK$ decay, with (a) and (b) for nonfactorizable annihilation,
 (c) and (d) for factorizable annihilation.}
 \label{fig1}
 \end{figure}

 The $B_{c}$ ${\to}$ $KK$ decay is the pure annihilation process.
 According to the effective Hamiltonian Eq.(\ref{eq:hamiltonian}),
 the lowest order Feynman diagrams are shown in FIG.\ref{fig1},
 where (a) and (b) are nonfactorizable topologies, (c) and (d) are
 factorizable topologies. After a straightforward calculation using
 the modified perturbative QCD formalism Eq.(\ref{eq:am01}), we
 find that the contributions of factorizable topologies are zero,
 which is a result of exact isospin symmetry. The decay amplitude
 comes only from the nonfactorizable topologies, and can be
 written as
 \begin{eqnarray}
 & & {\cal A}(B_{c}^{-}{\to}K^{-}K^{0}) \nonumber \\
 &=&-i\frac{G_{F}8{\pi}C_{F}f_{B_{c}}m_{B_{c}}^{4}}{\sqrt{2}N_{c}}
  V_{cb}V_{ud}^{\ast}{\int}_{0}^{1}{\bf d}x_{1}{\bf d}x_{2}{\bf d}x_{3}
 {\int}_{0}^{\infty}b_{1}{\bf d}b_{1}{\int}_{0}^{\infty}b_{2}{\bf d}b_{2}\,
 {\phi}_{B_{c}}(x_{1}) \nonumber \\ &{\times}& \Big\{
 {\alpha}_{s}(t_{a})C_{2}(t_{a})E(t_{a})H({\Delta},{\alpha},b_{1},b_{2})
  \Big[ {\phi}_{K^{-}}^{a}{\phi}_{K^{0}}^{a}\left(r_{b}+x_{1}-x_{3}\right)
  \nonumber \\ & &~~~~~~~~~~~~~~~~~~~~~~+
   r_{K^{-}}r_{K^{0}}{\phi}_{K^{-}}^{p}{\phi}_{K^{0}}^{p}
  \left( 4r_{b}+2x_{1}-x_{2}-x_{3}\right)
  \nonumber \\ & &~~~~~~~~~~~~~~~~~~~~~~+
   r_{K^{-}}r_{K^{0}} \left( {\phi}_{K^{-}}^{t}{\phi}_{K^{0}}^{p}
  +{\phi}_{K^{-}}^{p}{\phi}_{K^{0}}^{t} \right) \left(x_{3}-x_{2}\right)
  \nonumber \\ & &~~~~~~~~~~~~~~~~~~~~~~+
   r_{K^{-}}r_{K^{0}} {\phi}_{K^{-}}^{t}{\phi}_{K^{0}}^{t}
  \left(2x_{1}-x_{2}-x_{3}\right) \Big]
  \nonumber \\ &+&
 {\alpha}_{s}(t_{b})C_{2}(t_{b})E(t_{b})H({\Delta},{\beta},b_{1},b_{2})
  \Big[ {\phi}_{K^{-}}^{a}{\phi}_{K^{0}}^{a}\left(x_{2}-\bar{x}_{1}-r_{c}\right)
  \nonumber \\ & &~~~~~~~~~~~~~~~~~~~~~~+
   r_{K^{-}}r_{K^{0}}{\phi}_{K^{-}}^{p}{\phi}_{K^{0}}^{p}
  \left(x_{2}+x_{3}-2\bar{x}_{1}-4r_{c}\right)
  \nonumber \\ & &~~~~~~~~~~~~~~~~~~~~~~+
   r_{K^{-}}r_{K^{0}} \left( {\phi}_{K^{-}}^{t}{\phi}_{K^{0}}^{p}
  +{\phi}_{K^{-}}^{p}{\phi}_{K^{0}}^{t} \right) \left(x_{3}-x_{2}\right)
  \nonumber \\ & &~~~~~~~~~~~~~~~~~~~~~~+
   r_{K^{-}}r_{K^{0}} {\phi}_{K^{-}}^{t}{\phi}_{K^{0}}^{t}
   \left(x_{2}+x_{3}-2\bar{x}_{1}\right) \Big] \Big\}_{b_{2}=b_{3}}
 \label{eq:amplitude}
 \end{eqnarray}
 where the CKM matrix elements $V_{cb}V_{ud}^{\ast}$ $=$
 $A{\lambda}^{2}(1-{\lambda}^{2}/2-{\lambda}^{4}/8)$ $+$
 ${\cal O}({\lambda}^{8})$ with the phenomenological Wolfenstein
 parameterization.
 $r_{b}$ $=$ $m_{b}/m_{B_{c}}$ and $r_{c}$ $=$ $m_{c}/m_{B_{c}}$
 are the ratios of the mass of $b$ and $c$ quark to the mass of $B_{c}$
 mesons, respectively.
 $r_{K}$ $=$ ${\mu}_{K}/m_{B_{c}}$ $=$ $m_{K}^{2}/[m_{B_{c}}(m_{s}+m_{q})]$.
 $C_{F}$ $=$ $4/3$ is the $SU(3)$ color factor. $t_{a(b)}$ is the
 characteristic scale. ${\Delta}$ is the virtualities of internal
 gluons, which is a timelike variable for the pure annihilation $B_{c}$
 ${\to}$ $KK$ decay concerned. ${\alpha}$ and ${\beta}$ are the
 virtualities of internal quarks. $E$ and $H$ are the Sudakov factor
 and the hard kernel functions, respectively. Their expressions are
 listed in Appendix \ref{app02}.

 \section{Numerical results and discussions}
 \label{sec3}
 The branching ratio in the $B_{c}$ meson rest frame can be written as:
 \begin{equation}
  {\cal BR}(B_{c}{\to}KK)=\frac{{\tau}_{B_{c}}}{8{\pi}}
   \frac{p}{m_{B_{c}}^{2}}
  {\vert}{\cal A}(B_{c}{\to}KK){\vert}^{2}
   \label{eq:br-01},
 \end{equation}
 where $p$ is the center-of-mass momentum of $K$ mesons.
 The lifetime and mass of the $B_{c}$ meson are
 $m_{B_{c}}$ $=$ $6.276$ ${\pm}$ $0.004$ GeV and
 ${\tau}_{B_{c}}$ $=$ $0.46{\pm}0.07$ ps \cite{pdg2008},
 respectively.
 Other input parameters are
 \[ \begin{array}{lll}
   m_{c}=1.27^{+0.07}_{-0.11}~\hbox{\rm GeV}~\hbox{\cite{pdg2008}}, &
   {\lambda}=0.2257^{+0.0009}_{-0.0010}~\hbox{\cite{pdg2008}}, &
   f_{B_{c}}=489{\pm}4~\hbox{\rm MeV~\cite{pos180}}, \\
   m_{b}=4.20^{+0.17}_{-0.07}~\hbox{\rm GeV~\cite{pdg2008}}, &
   A=0.814^{+0.021}_{-0.022}~\hbox{\cite{pdg2008}}, &
   f_{K}=159.8{\pm}1.4{\pm}0.44~\hbox{\rm MeV~\cite{pdg2006}}.
   \end{array} \]
 If not specified explicitly, we shall take their central values as
 the default input. The numerical result of the branching ratio is
 \[ {\cal BR}(B_{c}^{-}{\to}K^{-}K^{0}){\approx}
    [1.63^{+0.67}_{-0.17}(m_{b})^{+0.35}_{-0.10}(m_{c})]{\times}
    [1{\pm}0.3\%(\hbox{CKM}){\pm}1.6\%(f_{B_{c}}){\pm}3.7\%(f_{K})]
    {\times}10^{-7}, \]
 where the errors come from the uncertainties of quark masses $m_{b}$
 and $m_{c}$, the CKM factor $V_{cb}V_{ud}^{\ast}$, and the decay
 constants $f_{B_{c}}$ and $f_{K}$. The largest error arises from
 the parameter of $m_{b}$, which can reach $40\%$. The errors
 arising from both the CKM factor and the decay constants are
 relatively small. Of course, there are some other uncertainties
 not considered here, such as the radiative corrections to the LCDAs
 of $B_{c}$ mesons, the final states interactions, etc. So
 the results might just be an estimation of the pQCD approach.

 Our estimation of the branching ratio ${\cal BR}(B_{c}{\to}KK)$
 is slightly different with the result in \cite{09121163},
 although they are calculated with the same pQCD approach
 resulting in the same order of magnitude ${\cal O}(10^{-7})$.
 Besides the input parameters, the reasons may be
 (1) whether the final states are on-mass-shell or not, and
 (2) whether the contributions of factorizable topologies are
 zero or not\footnotemark[3].
 \footnotetext[3]{Because of almost equal masses of the final states,
 the same two-particle LCDAs for the charged and neutral
 $K$ mesons are taken in our calculation. With this approximation,
 a similar conclusion, that the contributions of factorizable topologies
 cancel each other because of the isospin symmetry, can be found
 in $B_{s}$ ${\to}$ ${\pi}{\pi}$ decays with the pQCD approach
 \cite{prd70p034009}.}
 With appropriate input parameters, the results in \cite{09121163}
 and ours are in agreement with each other within an error range.

 As the arguments in \cite{prd80p114031}, the inconsistencies
 among various estimations of the branching ratio
 ${\cal BR}(B_{c}{\to}KK)$, such as ${\cal O}(10^{-6})$
 based on $B_{d}$ annihilation by using the relations among the
 charmless weak annihilation $B_{c}$ decay channels relying on
 the $SU(3)$ flavor symmetry \cite{prd80p114031},
 ${\cal O}(10^{-7})$ (or ${\cal O}(10^{-8})$ \cite{prd80p114031})
 based on perturbative one-gluon exchange with the pQCD (or QCDF)
 approach, arise from conceptually different methods.
 Anyway, for weak annihilation to light quarks in the final state, the
 tree annihilation $B_{c}^{-}$ ${\to}$ $d\bar{u}$ process is helicity
 suppressed because of small light quark masses, so that gluon
 emission either from the initial or final state must occur in
 this annihilation and the decay amplitude is then  $O({\alpha}_{s})$
 as given in pQCD. Both the estimations in \cite{prd80p114031,09121163}
 and our result are in accordance with an intuitive expectation for
 nonleptonic charmless $W$-annihilation of heavy meson decays
 which are usually suppressed.
 There are some additional factors for the tiny
 estimation of ${\cal BR}(B_{c}{\to}KK)$. One the is that although the
 $B_{c}{\to}KK$ decay is a tree weak annihilation process,
 its amplitude is color suppressed and associated with $C_{2}/N_{c}$.
 Another is that there is a large destructive interference between the
 nonfactorizable topologies due to the near equal final state particle
 masses. This can be clearly found in  Eq.(\ref{eq:amplitude}).
 The numerical results also confirm the cancellation between the
 nonfactorizable topologies, and give the strong phases ${\sim}$
 $-31^{\circ}$ and  ${\sim}$ $+127^{\circ}$ for FIG.\ref{fig1} (a)
 and (b), respectively.

 If the pQCD prediction is right, then there should be some $10^{3}$
 events for $B_{c}{\to}KK$ decay per year at the LHC.
 Considering the detection efficiency and selection efficiency, there
 would be just a few events per year. The signal of the pure weak
 annihilation $B_{c}{\to}KK$ decay would be very tiny at the LHC.
 As $B$ nonleptonic charmless decays, the charmless pure weak
 annihilation is expected to be small in $B_{c}$ nonleptonic decays,
 so the LHC measurement could confirm our understanding of the
 annihilation  terms in weak decays based on perturbative QCD.

 \section{Summary}
 \label{sec4}
 In this paper, we study the $B_{c}$ ${\to}$ $KK$ decay with the pQCD
 approach, which would call for another reassessment of the weak annihilation
 processes and might provide some valuable hints of our understanding on
 perturbative QCD and long-distance contributions.
 It is found that the contributions
 of factorizable annihilation topologies are zero, and that there is a large
 cancellation between the nonfactorizable topologies, which result
 in the branching ratio ${\cal BR}(B_{c}{\to}KK)$ ${\sim}$
 ${\cal O}(10^{-7})$.
 The branching ratio with the pQCD approach is so tiny that the
 $B_{c}$ ${\to}$ $KK$ decay might not be measured at the LHC experiments.

 \section*{Acknowledgments}
 This work is supported by both National Natural Science Foundation
 of China (under Grant No. 10805014) and the program for Science \&
 Technology Innovation Talents in Universities of Henan Province,
 China (under Grant No. 2010HASTIT001). We would like to thank the
 referees for their helpful comments.

 \begin{appendix}
 \section{Distribution amplitude of the $K$ meson}
 \label{app01}
 The expression of the LCDAs of the $K$ meson incluing higher-twist
 contributions can be found in \cite{jhep0605004}. In our calculation,
 the twist-2 distribution amplitude ${\phi}_{K}^{a}$ and the twist-3
 distribution amplitude ${\phi}_{K}^{p}$ and ${\phi}_{K}^{t}$
 are \cite{klcda}
  \begin{eqnarray}
  \hbox{twist-2} &&
 {\phi}_{K}^{a}(x)\,=\,\frac{f_{K}}{2\sqrt{2N_{c}}}6x\bar{x}
  \Big\{1+0.17C_{1}^{3/2}(t)+0.115C_{2}^{3/2}(t)\Big\} \\
  \hbox{twist-3} &&
 {\phi}_{K}^{p}(x)\,=\,\frac{f_{K}}{2\sqrt{2N_{c}}}
  \Big\{1+0.24C_{2}^{1/2}(t)-0.12C_{4}^{1/2}(t)\Big\}, \\
  \hbox{twist-3} &&
 {\phi}_{K}^{t}(x)\,=\,\frac{-f_{K}}{2\sqrt{2N_{c}}}
  \Big\{C_{1}^{1/2}(t)+0.35C_{3}^{1/2}(t)\Big\}
 \end{eqnarray}
 where the decay constant $f_{K}$ $=$ $160$ MeV.
 $t$ $=$ $x$ $-$ $\bar{x}$ $=$ $2x$ $-$ $1$.
 The Gegenbauer polynomials are
 \[ \begin{array}{lll}
    \displaystyle C_{1}^{3/2}(z)=3z, &~~
  & \displaystyle C_{2}^{3/2}(z)=\frac{3}{2}(5z^{2}-1), \\
    \displaystyle C_{1}^{1/2}(z)=z, &
  & \displaystyle C_{2}^{1/2}(z)=\frac{1}{2}(3z^{2}-1), \\
    \displaystyle C_{3}^{1/2}(z)=\frac{1}{2}(5z^{3}-3z), &
  & \displaystyle C_{4}^{1/2}(z)=\frac{1}{8}(35z^{4}-30z^{2}+3)
 \end{array} \]

 \section{Some parameters and formulas}
 \label{app02}
 The expression of the Sudakov factors $E$ is
 \begin{equation}
 E(t)\,=\,{\exp}\left(-S_{B_{c}}(t)-S_{K^{-}}(t)-S_{K^{0}}(t)\right)
 \end{equation}
 where
 \begin{eqnarray}
 S_{B_{c}}(t)&=&s(x_{1}p_{1}^{+},b_{1})
   +2{\int}_{1/b_{1}}^{t}\frac{{\bf d}{\mu}}{\mu}{\gamma}_{q} \\
 S_{K^{-}}(t)&=&s(x_{2}p_{2}^{+},b_{2})+s(\bar{x}_{2}p_{2}^{+},b_{2})
   +2{\int}_{1/b_{2}}^{t}\frac{{\bf d}{\mu}}{\mu}{\gamma}_{q} \\
 S_{K^{0}}(t)&=&s(x_{3}p_{3}^{-},b_{3})+s(\bar{x}_{3}p_{3}^{-},b_{3})
   +2{\int}_{1/b_{3}}^{t}\frac{{\bf d}{\mu}}{\mu}{\gamma}_{q}
 \end{eqnarray}
 The anomalous dimension of the quark is ${\gamma}_{q}$ $=$
 $-{\alpha}_{s}/{\pi}$. The explicit expression of $s(Q,b)$
 can be found in \cite{npb642p263}.

 The hard kernel function $H$ is defined as follows
 \begin{eqnarray}
 H({\Delta},Z,b_{1},b_{2})&=&\left\{{\theta}(b_{1}-b_{2})
   \frac{i{\pi}}{2}H_{0}^{(1)}(\sqrt{\Delta}b_{1})J_{0}(\sqrt{\Delta}b_{2})
  +\left(b_{1}{\leftrightarrow}b_{2}\right)\right\}
   \nonumber \\ &{\times}&
   \left\{ {\theta}(Z)K_{0}(\sqrt{Z}b_{1})+{\theta}(-Z)
   \frac{i{\pi}}{2}H_{0}^{(1)}(\sqrt{{\vert}Z{\vert}}b_{1}) \right\}
 \end{eqnarray}
 where the hard scales are
  \begin{eqnarray}
   {\Delta}&=&m^{2}(1-x_{2})(1-x_{3}) \\
  -{\alpha}&=&m^{2}(x_{1}-x_{2})(x_{1}-x_{3})-m_{b}^{2} \\
  -{\beta} &=&m^{2}(x_{1}+x_{2}-1)(x_{1}+x_{3}-1)-m_{c}^{2} \\
  t_{a}&=&\max\left(\sqrt{\Delta},\sqrt{\alpha},1/b_{1},1/b_{2}\right) \\
  t_{b}&=&\max\left(\sqrt{\Delta},\sqrt{\beta},1/b_{1},1/b_{2}\right)
  \end{eqnarray}

 \end{appendix}

 \end{document}